\documentclass[twoside,twocolumn,9pt]{article}  %% default paper size is letter, bizarrely?
\usepackage{extsizes}
\usepackage[super,sort&compress,comma]{natbib}
\usepackage[version=3]{mhchem}
\usepackage[left=1.5cm, right=1.5cm, top=1.785cm, bottom=2.0cm]{geometry}
\usepackage{balance}
\usepackage{widetext}
\usepackage{sectsty}
\usepackage{graphicx}
\usepackage{lastpage}
\usepackage[format=plain,justification=raggedright,singlelinecheck=false,font={stretch=1.125,small,sf},labelfont=bf,labelsep=space]{caption}

\usepackage{float}
\usepackage{fancyhdr}
\usepackage{fnpos}
\usepackage[english]{babel}
\usepackage{array}
\usepackage{droidsans}
\usepackage{charter}
\usepackage[T1]{fontenc}
\usepackage[usenames,dvipsnames]{xcolor}
\usepackage{setspace}
\usepackage[compact]{titlesec}
\definecolor{cream}{RGB}{222,217,201}

\usepackage{cmap}
\usepackage{fixmath}

\usepackage[slantedGreek]{mathptmx}
\usepackage{times}
\usepackage{fixmath}
\usepackage[Symbolsmallscale]{upgreek}
\usepackage[italic]{mathastext}

\usepackage{mciteplus}
\usepackage{mleftright}

\usepackage{booktabs}

\usepackage[pdftex, bookmarks, pdffitwindow=false, pdfstartview=FitH, pdfdisplaydoctitle, colorlinks, plainpages=false,pdftitle={Investigating the role of boundary bricks in DNA brick self-assembly}, pdfauthor={Hannah Wayment-Steele, Daan Frenkel, Aleks Reinhardt}, pdfpagelabels, hypertexnames, citecolor={blue},linkcolor={red}, urlcolor={violet}, pdflang={en}, hyperfootnotes=false, breaklinks]{hyperref}

\usepackage{amsmath}

\usepackage{flafter}
\usepackage{setspace}

\usepackage[stretch=15,shrink=15,step=1]{microtype}
\usepackage[obeyall]{siunitx}

\newcommand{\refSub}[2]{\hyperref[#2]{\ref{#2}(#1)}}
\newcommand{\refSubNB}[2]{\hyperref[#2]{\ref{#2}#1}}
\newcommand{\refSubAlt}[2]{\hyperref[#2]{(#1)}}

\begin{document}

\pagestyle{fancy}
\thispagestyle{plain}
\fancypagestyle{plain}{

%%%HEADER%%%
\fancyhead[C]{\includegraphics[width=18.5cm]{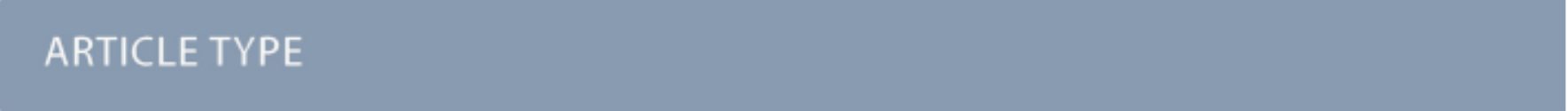}}
\fancyhead[L]{\hspace{0cm}\vspace{1.5cm}}%\includegraphics[height=30pt]{head_foot/journal_name}}
\fancyhead[R]{\hspace{0cm}\vspace{1.7cm}\includegraphics[height=55pt]{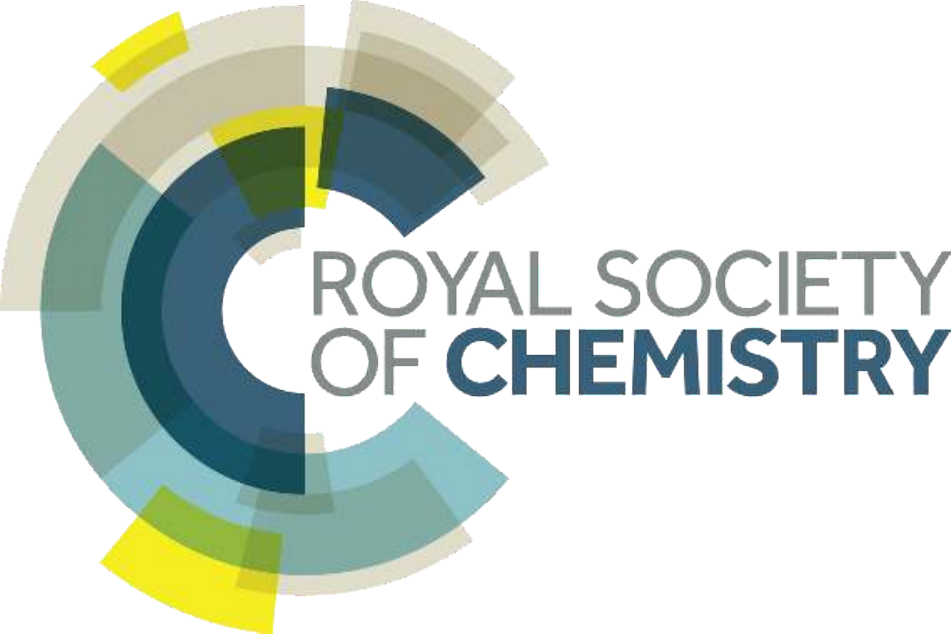}}
\renewcommand{\headrulewidth}{0pt}
}
%%%END OF HEADER%%%

%%%PAGE SETUP - Please do not change any commands within this section%%%
\makeFNbottom
\makeatletter
\renewcommand\LARGE{\@setfontsize\LARGE{15pt}{17}}
\renewcommand\Large{\@setfontsize\Large{12pt}{14}}
\renewcommand\large{\@setfontsize\large{10pt}{12}}
\renewcommand\footnotesize{\@setfontsize\footnotesize{7pt}{10}}
\makeatother

\renewcommand{\thefootnote}{\fnsymbol{footnote}}
\renewcommand\footnoterule{\vspace*{1pt}%
\color{cream}\hrule width 3.5in height 0.4pt \color{black}\vspace*{5pt}}
\setcounter{secnumdepth}{5}

\makeatletter
\renewcommand\@biblabel[1]{#1}
\renewcommand\@makefntext[1]%
{\noindent\makebox[0pt][r]{\@thefnmark\,}#1}
\makeatother
\renewcommand{\figurename}{\small{Fig.}~}
\sectionfont{\sffamily\Large}
\subsectionfont{\normalsize}
\subsubsectionfont{\bf}
\setstretch{1.125} %In particular, please do not alter this line.
\setlength{\skip\footins}{0.8cm}
\setlength{\footnotesep}{0.25cm}
\setlength{\jot}{10pt}
\titlespacing*{\section}{0pt}{4pt}{4pt}
\titlespacing*{\subsection}{0pt}{15pt}{1pt}
%%%END OF PAGE SETUP%%%

%%%FOOTER%%%
\fancyfoot{}
\fancyfoot[LO,RE]{\vspace{-7.1pt}\includegraphics[height=9pt]{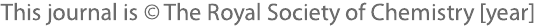}}
\fancyfoot[CO]{}%\vspace{-7.1pt}\hspace{13.2cm}\includegraphics{head_foot/RF}}
\fancyfoot[CE]{\vspace{-7.2pt}\hspace{-14.2cm}\includegraphics{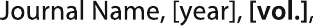}}
\fancyfoot[RO]{\footnotesize{\sffamily{1--\pageref{LastPage} ~\textbar  \hspace{2pt}\thepage}}}
\fancyfoot[LE]{\footnotesize{\sffamily{\thepage~\textbar\hspace{3.45cm} 1--\pageref{LastPage}}}}
\fancyhead{}
\renewcommand{\headrulewidth}{0pt}
\renewcommand{\footrulewidth}{0pt}
\setlength{\arrayrulewidth}{1pt}
\setlength{\columnsep}{6.5mm}
\setlength\bibsep{1pt}
%%%END OF FOOTER%%%

%%%FIGURE SETUP - please do not change any commands within this section%%%
\makeatletter
\newlength{\figrulesep}
\setlength{\figrulesep}{0.5\textfloatsep}

\newcommand{\topfigrule}{\vspace*{-1pt}%
\noindent{\color{cream}\rule[-\figrulesep]{\columnwidth}{1.5pt}} }

\newcommand{\botfigrule}{\vspace*{-2pt}%
\noindent{\color{cream}\rule[\figrulesep]{\columnwidth}{1.5pt}} }

\newcommand{\dblfigrule}{\vspace*{-1pt}%
\noindent{\color{cream}\rule[-\figrulesep]{\textwidth}{1.5pt}} }

\makeatother
%%%END OF FIGURE SETUP%%%

%%%TITLE, AUTHORS AND ABSTRACT%%%
\twocolumn[
  \begin{@twocolumnfalse}
\vspace{3cm}
\sffamily
\begin{tabular}{m{4.5cm} p{13.5cm} }

\href{https://doi.org/10.1039/c6sm02719a}{doi:10.1039/c6sm02719a} & \noindent\LARGE{\textbf{Investigating the role of boundary bricks in DNA brick self-assembly}} \\%Article title goes here instead of the text "This is the title"
\vspace{0.3cm} & \vspace{0.3cm} \\

  & \noindent\large{Hannah K.~Wayment-Steele,$^{a,\ddag}$ Daan Frenkel$^{a}$ and Aleks Reinhardt$^{a}$} \\%Author names go here instead of "Full name", etc.

\includegraphics{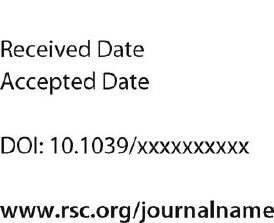} & \noindent\normalsize{In the standard DNA brick set-up, distinct 32-nucleotide strands of single-stranded DNA are each designed to bind specifically to four other such molecules. Experimentally, it has been demonstrated that the overall yield is increased if certain bricks which occur on the outer faces of target structures are merged with adjacent bricks. However, it is not well understood by what mechanism such `boundary bricks' increase the yield, as they likely influence both the nucleation process and the final stability of the target structure. Here, we use Monte Carlo simulations with a patchy particle model of DNA bricks to investigate the role of boundary bricks in the self-assembly of complex multicomponent target structures. We demonstrate that boundary bricks lower the free-energy barrier to nucleation and that boundary bricks on edges stabilize the final structure. However, boundary bricks are also more prone to aggregation, as they can stabilize partially assembled intermediates. We explore some design strategies that permit us to benefit from the stabilizing role of boundary bricks whilst minimizing their ability to hinder assembly; in particular, we show that maximizing the total number of boundary bricks is not an optimal strategy.} \\

\end{tabular}

 \end{@twocolumnfalse} \vspace{0.6cm}

  ]
%%%END OF TITLE, AUTHORS AND ABSTRACT%%%

%%%FONT SETUP - please do not change any commands within this section
\renewcommand*\rmdefault{bch}\normalfont\upshape
\rmfamily
\section*{}
\vspace{-1cm}

\footnotetext{\textit{$^{a}$~Department of Chemistry, University of Cambridge, Lensfield Road, Cambridge, CB2~1EW, United Kingdom.}}
\footnotetext{\textit{$\ddag$~Present address: Department of Chemistry, Stanford University, Stanford, California 94305, USA.}}

% \nocite{rsc-control} %% fake citation to use the rsc.bst options -- this allows us to print journal titles, which is important for the arXiv references

\section{Introduction}

\begin{figure*}
\centering
\includegraphics{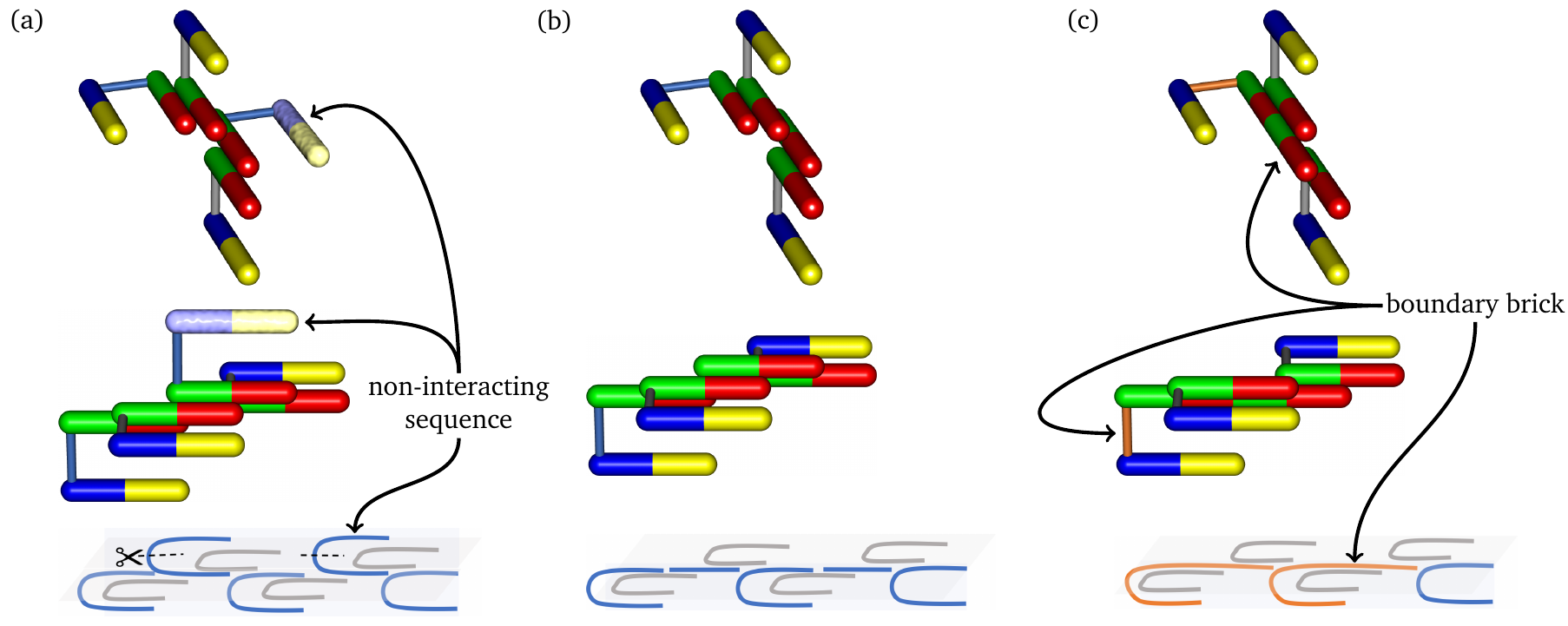}
  \caption{Schematic of the boundary brick set-up in two simplified representations. In the top representation, shown in two orientations, only a single surface molecule is shown alongside its neighbours; the remainder of the structure has been omitted. The cylinders represent single-stranded DNA molecules; cylinders that are adjacent to one another are hybridized, but are slightly off-set from one another for clarity. Each molecule has four domains, shown in red, green, blue and yellow. Red sections of the cylinders are bonded with green sections in the correctly assembled structure. (a) In the assembled structure, bricks are staggered in rows along one of the cartesian axes. Each of the four domains of a molecule hybridizes with a different neighbouring molecule. At surfaces, every other row has half a brick protruding from the surface . Since there are no neighbouring molecules on the surface, these parts of the DNA molecules have no neighbours in the target structure and should be non-interacting.  (b) If the non-interacting sequence is removed altogether by bisecting the surface bricks, a 16-nucleotide brick occurs in every other row. (c) Each 16-nucleotide strand can then be combined with the strand immediately preceding it to create 48-nucleotide strands known as boundary bricks.}
 \label{fig-BB-schematic}
\end{figure*}

Since their discovery,\cite{Wei2012, Ke2012} two-dimensional DNA tiles and three-dimensional DNA bricks have gathered interest as a completely modular DNA nanomaterial. In the DNA brick set-up, short, 32-nucleotide long single-stranded DNA molecules have sequences chosen such that they hybridize specifically with four other distinct single-stranded molecules. The interactions are chosen such that favourable bonding occurs when these molecules are arranged in a target structure. If a big cubic structure is designed in this way in the first instance, other structures can rapidly be designed using the same set of starting bricks by merely omitting a subset of the bricks.\cite{Ke2012} While DNA origami\cite{Seeman2003, Seeman1982, Winfree1998, Rothemund2006, Torring2011} is currently the most popular strategy for fabricating DNA nanomaterials, unlike with DNA bricks, DNA origami entails a long `scaffold' single-stranded DNA molecule which is linked with shorter `staple' molecules to fold the scaffold strand into the target shape, and designing a new target structure therefore requires starting from scratch with an entirely new set of staple strands.

DNA brick self-assembly is also perhaps the best example of a viable addressable\cite{Frenkel2015} self-assembled system: each subunit in the target structure is unique, and knowing the identity of a particle therefore means knowing its location, and vice versa. Systems with addressable complexity and their potential for designing structures with arbitrary shape and complexity have great promise in fields ranging from medical applications to nanoelectronics.\cite{Cademartiri2015}

A recent application of DNA bricks as a nano-breadboard for chromophore-based excitonic gates\cite{Cannon2015} exemplifies the benefits of DNA bricks over DNA origami. In excitonic devices, where the FRET radius is less than \SI{5}{\nano\metre}, it is necessary to have nanometre-scale control over the placement of chromophores. DNA bricks have an advantage over DNA origami by having twice the spatial resolution: it is difficult to functionalize the scaffold strand in DNA origami, and thus only the staple strands, one out of two strands in any helix, are available for functionalization.\cite{Cannon2015} By contrast, in DNA brick structures, all strands are available for functionalization. In such technologies, the excitonic transmission behaviour is challenging to predict and the modular nature of DNA bricks allows for straightforward modification of structures, permitting a number of possible layouts to be tested and screened rapidly.\cite{Cannon2015} Moreover, these benefits may prove useful for other applications as well, for instance in scaffolding for multi-enzyme complexes for single-molecule reactions,\cite{Fu2014} molecular rulers,\cite{Schmied2014} inorganic nanoparticle synthesis\cite{Sun2014} and nano-robots.\cite{Amir2014}

In DNA brick structures, the final structure is designed to be the thermodynamic product, a benefit over folding assembly structures, where it is difficult to predict if the designed target structure is the preferred equilibrium structure.\cite{Cademartiri2015,Snodin2016} However, DNA brick assembly has a much more complex pathway to assembly.\cite{Reinhardt2014, Jacobs2015b, Jacobs2016} Because there are a vast number of intermediate states that all have similar energies, DNA bricks are very prone to kinetic traps.\cite{Cademartiri2015, Whitelam2015} This is a disadvantage in comparison to folding assembly, where the constraint offered by having all interacting particles on a backbone offers more direction to the final assembled state. Experimentally, typical DNA brick yields range from a few per cent to 30 per cent,\cite{Ke2012} whereas yields for some DNA origami structures are approaching \SI{99}{\percent}.\cite{Wagenbauer2014,FootNoteYields} However, because we have control over interactions between DNA brick subunits, it should be possible to design interactions that can direct the assembly to avoid kinetic traps.

One design strategy implemented for increasing the yield in experiment was to include larger bricks at the surfaces of target structures.\cite{Ke2012} Because bricks are staggered in the $xz$ and $yz$ planes of the structures, at the faces of a structure, in alternating rows a brick must be bisected (Fig.~\refSub{a}{fig-BB-schematic}), leaving behind 16-nucleotide half-bricks in every other row (Fig.~\refSub{b}{fig-BB-schematic}). These half-bricks were then connected to the bricks in the row preceding them to form larger 48-nucleotide bricks, termed `boundary bricks' (BBs) (Fig.~\refSub{c}{fig-BB-schematic}). The use of BBs was shown to increase the yield by a factor of 1.4 and was implemented for all subsequent structures in the experiments of Ke~\textit{et al}.\cite{Ke2012}

Despite the effectiveness of BBs in increasing the yield, the cause of this observed effect has not been well studied. There are two principal mechanisms one can envisage by which BBs could lead to an increase of the yield. Firstly, because they are larger and have more interaction domains than regular bricks, they could serve as a larger seed particle to promote nucleation. Secondly, they may stabilize the final structure by binding the edge half-bricks that have fewer interaction points to bind to the rest of the structure.

To investigate the effects of BBs on the DNA brick nucleation and assembly process, in this work we extend a simulation-based model previously used with success to describe DNA bricks\cite{Reinhardt2014,Jacobs2015, Jacobs2015b, Reinhardt2016, Reinhardt2016b} to include BBs. With this model, we use Monte Carlo simulations to show that depending on the location, BBs may differ in their contribution to increasing the nucleation rate or stabilizing the final structure. We also demonstrate that BBs are more prone to aggregation than regular bricks, and we suggest a method to overcome this whilst still benefiting from the stabilizing effect of incorporating BBs.

\section{Methods}
\begin{figure}[t!]
\centering
\includegraphics{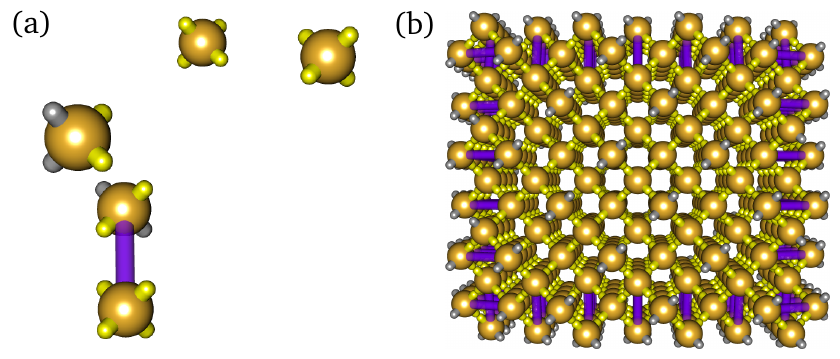}
\caption{(a) Example of one BB and several monomer bricks in solution, showing the tetrahedral patch arrangement. The rigid `bond' connecting the DNA bricks that constitute a single BB is shown in violet. (b) Example of one of the target structures with all boundary bricks shown. Interacting patches are shown in yellow, and non-interacting poly-T sequence patches are shown in grey.}
\label{fig-BB-ex-1}
\end{figure}
We perform Metropolis Monte Carlo\cite{Metropolis1953} simulations on a cubic lattice in the canonical ensemble with `virtual moves'\cite{Whitelam2007, *Whitelam2008} accounting for the motion of clusters. To be able to probe the time and length scales needed to observe assembly behaviour in a computational context, we model DNA bricks as spheres with four `sticky' interacting patches, representing the four 8-nucleotide sequence domains of DNA bricks, placed equidistantly on the sphere's surface to form a tetrahedral shape.\cite{Reinhardt2014} The dihedral angles in the DNA brick structures are roughly \ang{90}, which means that the centres of mass of each DNA brick in the final structure form a distorted diamond lattice,\cite{Ke2012} so describing each brick as a tetrahedron serves as a reasonable first-order approximation of the experimental geometry.\cite{Reinhardt2014}

\begin{figure}[tbp]
\centering
\includegraphics{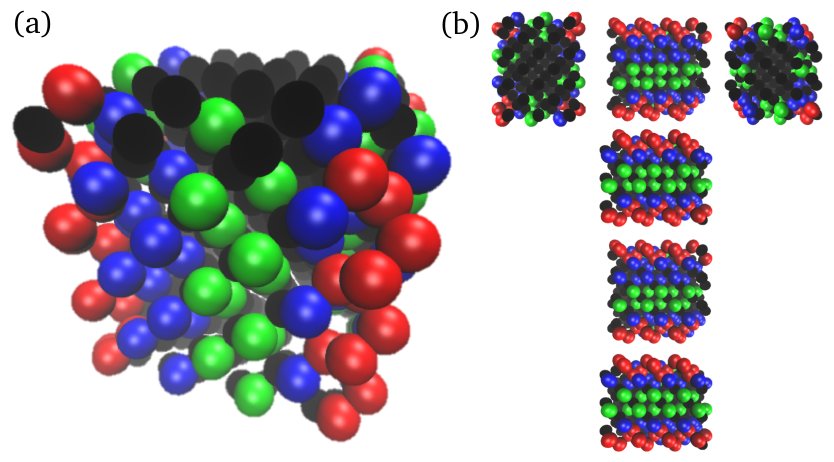}
\caption{(a) Corner view of 374-particle structure with 84 boundary bricks on faces in $xz$ and $yz$ planes. Edge BBs are shown in red, central face BBs are shown in green, and remaining face BBs are shown in blue. Monomer bricks are shown in black. (b) Net projection of the structure. }
\label{fig-BBs-allTypes}
\end{figure}

In an extension of the previous model,\cite{Reinhardt2014} we model boundary bricks as dimers of these particles, i.e.~as two patchy particles that are connected by a rigid bond of length corresponding to their distance in the target structure (Fig.~\ref{fig-BB-ex-1}), with (at least) two patches on one of the particles given a poly-T sequence\cite{FootNotePolyT} to passivate them (cf.~Fig.~\refSub{a}{fig-BB-schematic}).\cite{FootNoteModelRemark} The two particles connected in this way are also fixed in their orientation with respect to one another.\cite{FootNoteBoundaryCount}

To investigate the effect of BBs on structure nucleation and final stability, we perform simulations of a DNA brick structure with 374 bricks in the canonical ensemble (i.e.~at a constant number of particles, volume and temperature).  In the majority of simulations reported here, the density of each type of brick was set to $1/(62a)^3$, where $a$ is the lattice grid parameter, and a single copy of each brick that appears in the target system was present in the simulation box. The simulated structures contain up to 84 boundary bricks on the faces in the $xz$ and $yz$ planes (Fig.~\ref{fig-BBs-allTypes}), as described below. Simulations were run for structures with DNA-specific interactions between bricks. For each set of bonded patches in the final target structure, complementary sequences were randomly generated and assigned. The interaction strength of hybridization between two complementary DNA strands largely depends on the proportion of paired guanine/cytosine (GC) pairs in the sequence. The average GC content for the structure studied was \SI{44.6}{\percent}, with standard deviation \SI{14.2}{\percent}. Each patch can interact with every other patch provided the patches point at each other and the corresponding particles are diagonally adjacent to each other, and the energy of interaction corresponds to the hybridization free energy obtained from the SantaLucia thermodynamic model.\cite{SantaLucia2004}

In simulations investigating whether BBs in different locations of the final structure had differing impacts on structure nucleation and final stability, we chose to include various sets of BBs out of the total possible 84 BBs for this structure. One subset consisted of 26 BBs on the edges of the cubic structure, henceforth referred to as the `edge-BB structure', shown in red in Fig.~\ref{fig-BBs-allTypes}. The other subset included 26 BBs at the centre of the faces of the cubic structure, henceforth referred to as the `face-BB structure', with BBs shown in green in Fig.~\ref{fig-BBs-allTypes}. The structure with all possible BBs\cite{FootNoteFaceBBs} (red, green and blue) will be referred to as the `all-BB structure'. In the original experimental design, bricks with sequence domains on the external faces of the structure had those domains either removed or replaced with non-interacting poly-T sequences.\cite{Ke2012} Our model similarly passivates patches of bricks that are on the external face by assigning them a poly-T sequence. We ensured that there is no significant difference in the number of passivated patches on BBs between the edge and face conditions.

In addition to simulations of structures with specific DNA interactions, we have also run simulations of structures with designed interactions all having the same fixed interaction energy. This simplification still retains the specificity of interactions required for addressable self-assembly, since patches still have a unique identity and only interact with specific patches with which they were designed to bond in the target structure, but removes variation in interaction strengths that arises from DNA sequence dependence. In such simulations, all designed interactions were assigned a fixed interaction energy of $\varepsilon/k_\text{B} =-\SI{4000}{\kelvin}$, which corresponds roughly to an average sequence interaction strength at \SI{320}{\kelvin}.\cite{Jacobs2015b} All other (`incidental') patch--patch interactions were set to zero.

\section{Results and discussion}

\begin{figure*}[t!]
\centering
\includegraphics{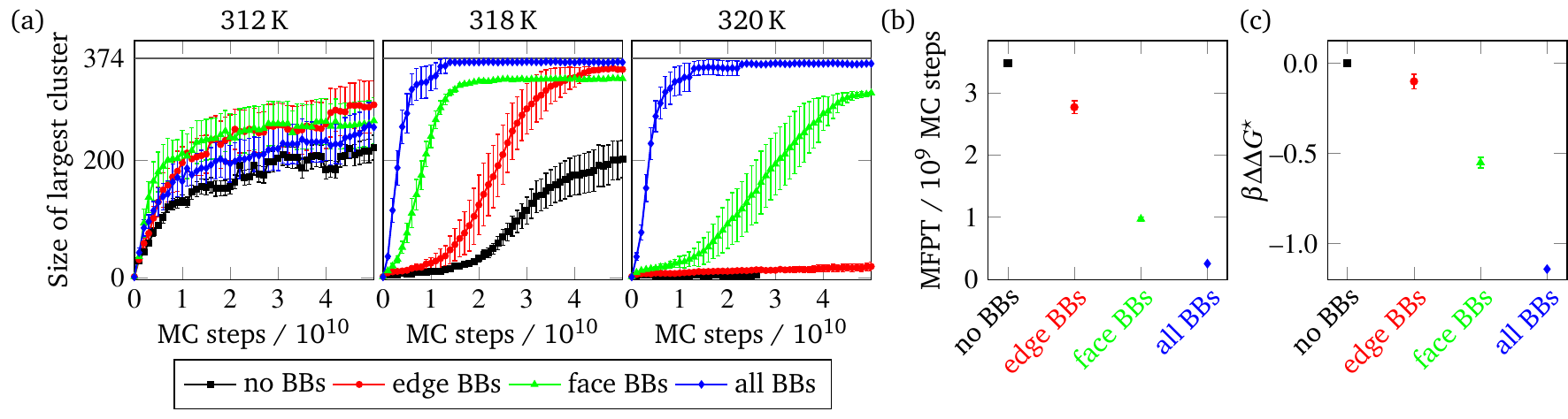}
\caption{(a) Assembly of structures with varying subsets of BBs as a function of simulation time for a range of temperatures, as indicated. SantaLucia parameters were used for interactions between patches. The target structure comprises 374 particles; this size is indicated by a grey line. (b) The mean first-passage time (MFPT) for clusters in the given structures to reach size 8, at \SI{318}{\kelvin}. (c) Calculated reduction in the nucleation free-energy barrier height based on the MFPT of the structures. Results were averaged over 15 independent simulations in (a) and 60 independent simulations in (b) and (c). Error bars represent the standard deviation.}
\label{fig-nuc-rate}
\end{figure*}

\subsection{Effect of BBs on the nucleation rate}

We ran brute-force Monte Carlo (MC) simulations of the self-assembly of target structures from a vapour of monomers (corresponding in reality to a dilute solution of monomers) at a range of fixed temperatures to observe the dependence of assembly behaviour on temperature, since the self-assembly process of DNA brick systems has been shown to be particularly sensitive to temperature.\cite{Reinhardt2014, Jacobs2015b} We have run such simulations on four sets of building blocks, namely the edge-, face- and all-BB structures as defined above and a system with no boundary bricks. We show the size of the largest correctly assembled cluster in the system as a function of Monte Carlo time in Fig.~\refSub{a}{fig-nuc-rate}. We find that, for the structure considered here, at temperatures below approximately \SI{315}{\kelvin}, assembly is dominated by unintended aggregation. Because lower temperatures favour both correct bonding and incorrect bonding, and there are statistically many more ways to bond incorrectly, the structure nucleates and assembles to some degree but quickly becomes kinetically trapped in a misassembled state and is then unable to assemble correctly any further. Optimal assembly is observed between about \SI{317}{\kelvin} and \SI{318}{\kelvin}. In this range, the all-BB structure grows the most rapidly, followed by the face-BB structure and then the edge-BB structure. The no-BB structure takes the longest to nucleate and grow. The same trend was observed in structures with fixed designed interactions (FDI), confirming that this was not an artefact of any possible difference in GC content of BBs in the edge and face structures. At \SI{319}{\kelvin} and above, the edge-BB and no-BB structures take significantly longer to nucleate, but the all-BB structure largely assembles up to about \SI{326}{\kelvin}, since the large number of BBs makes the bonding much more favourable for this system.

We can estimate the effect of BBs on increasing the nucleation rate by measuring the mean first-passage time (MFPT). This approach is commonly used when computing nucleation rates in molecular dynamics simulations;\cite{Wedekind2007b, *Nicholson2016} the mean first-passage time corresponds to the time needed on average for a stochastic process to reach a certain state for the first time. Although MC simulations do not faithfully reproduce the various time scales that may be involved in a dynamic process, we can nevertheless estimate the relative effect of BBs on the nucleation rate by calculating the MFPT for the different BB structures, using MC steps as a time step approximation.

In order to compute the MFPT for nucleation in our simulations, we recorded the number of MC steps required for the size of the largest correctly assembled cluster in the system to reach 8 particles and large-scale growth to begin. This cluster size was chosen since we have previously shown that the critical cluster at temperatures at which DNA bricks can nucleate usually comprises 8 bricks,\cite{Jacobs2015b} and the largest cluster in the system typically grows rather than shrinks once clusters grow beyond this size. The reason for this well-defined critical cluster size is that 9 bricks are required to form a subunit with two closed `cycles' (i.e.~closed loops of particles that are bonded to one another), and, as a monomer comes in to close a cycle, two bonds are formed simultaneously, energetically stabilizing the resulting structure at roughly the same entropic cost. The critical cluster structure comprises one brick less than this stabilized bicyclic motif.\cite{Jacobs2015b}

The difference in the free-energy barrier height relative to the system with no boundary bricks, $\upDelta\upDelta G^{\star} \equiv \upDelta G^{\star}-\upDelta G^{\star}_\text{no-BB}$, is calculated from the ratios of the average nucleation rate $R$, which is the reciprocal of the MFPT. We use the classical nucleation theory relation\cite{Turnbull1949,*Auer2001,*Auer2002}
\begin{equation}
R=N_\text{S} Z j \exp \mleft( -\beta \upDelta G^{\star} \mright),
\end{equation}
where $N_\text{S}$ is the number of nucleation sites, $Z$ is the Zeldovich factor,\cite{Zeldovich1943} $j$ is the rate at which molecules attach to the nucleus, and $\upDelta G^{\star}$ is the free energy required to self-assemble the critical nucleus from a dilute solution. Although we do not know $N_\text{S}$, $Z$ or $j$ for this system, we assume that they are roughly the same for all systems, regardless of the number of BBs present, particularly as the dependence on the nucleation free-energy barrier is exponential and the remaining terms are not. If we take the ratio of nucleation rates, these terms thus (approximately) cancel out, giving a ratio of
\begin{equation}
\frac{R}{R_\text{no-BB}} =\exp \mleft( -\beta \upDelta \upDelta G^{\star} \mright).
\end{equation}
The mean first-passage times and the corresponding values of $\upDelta \upDelta G^{\star}$ are shown in Fig.~\refSub{b}{fig-nuc-rate} and \refSubAlt{c}{fig-nuc-rate}.

The relative changes in the free-energy barriers are in agreement with the trends observed qualitatively from monitoring the largest cluster over time, with the edge BBs having the smallest reduction in the free-energy barrier, followed by face BBs, and finally followed by the system with all possible BBs. Although the latter system has 3.2 times as many boundary bricks as do the edge and face-BB systems, there is only a relatively small decrease in the free-energy barrier from the edge and face-BB systems to the all-BB system. This is perhaps not particularly surprising, since the reduction of the free-energy barrier is affected principally by the bricks first involved in nucleation, not their overall number.

\begin{figure*}[!tb]
\centering
\includegraphics{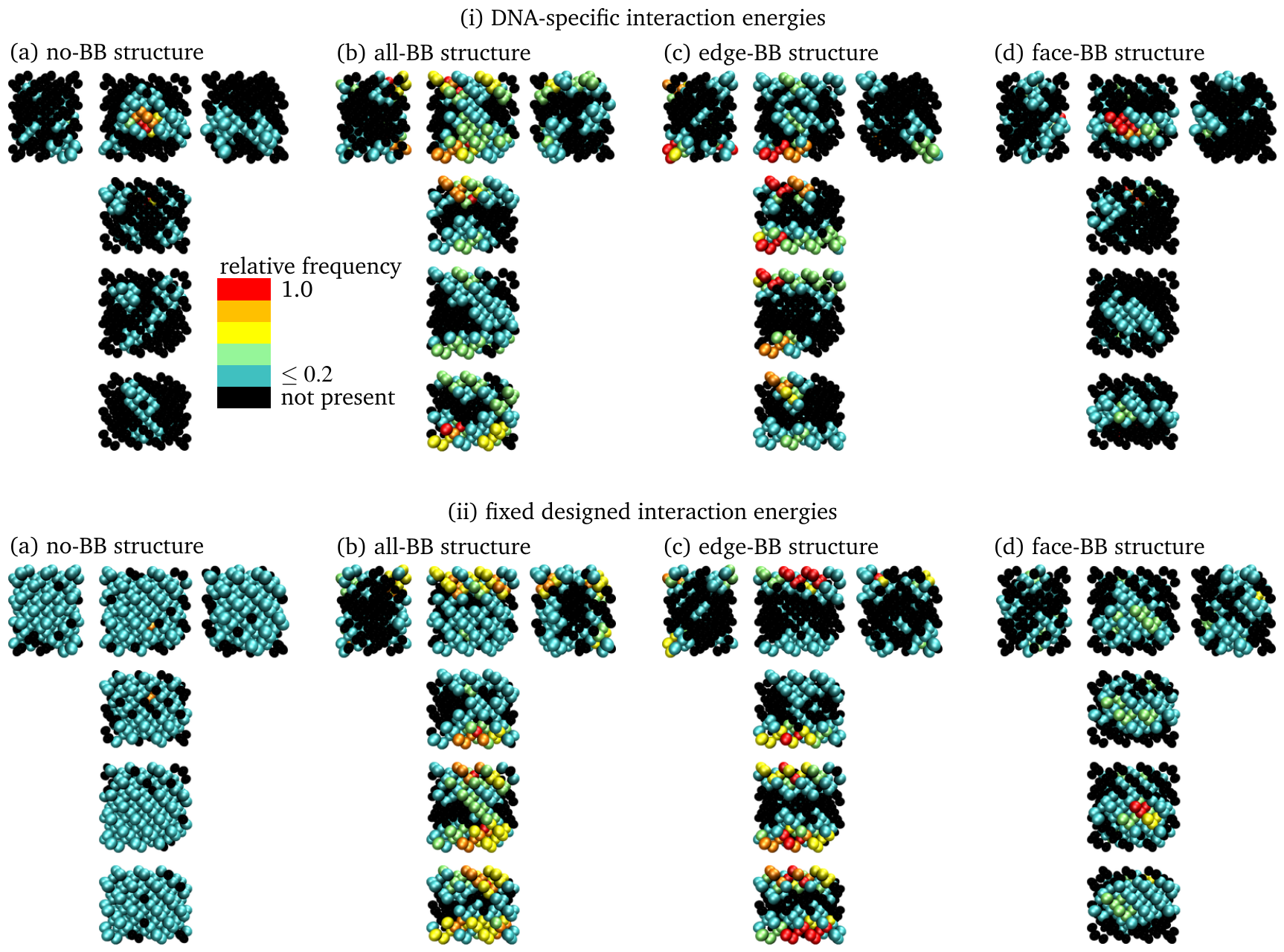}
\caption{Relative frequencies of bricks in initial nucleation clusters of size 9 for structures with (i) DNA-specific (SantaLucia) interaction energies and (ii) fixed designed interaction energies, overlaid onto the net of the target structure to showcase the location of the bricks involved in nucleation. Frequencies averaged over 60 independent simulations for each structure.}
\label{fig-BBs-nucleationSites}
\end{figure*}

The difference in the MFPT between the face-BB structures and edge-BB structures is interesting, as in many ways besides the obvious difference in the location of the BBs, the structures are identical. Both have 26 BBs, and they have no significant difference in the number of interacting patches or GC content. The same trend is also observed in FDI simulations, in which all designed interactions have a fixed interaction energy, indicating this is not an effect of GC content in the structures. The increased nucleation rate in face-BB structures likely arises because there are more interactions between face BBs and non-face bricks than there are interactions between edge BBs and non-edge bricks, since the latter have fewer neighbouring molecules.

\subsection{Nucleation location}
In order to understand better how BBs are involved in initial nucleation and growth, we have identified which bricks are involved in the nucleated clusters for the structures studied. We chose to investigate clusters comprising 9 monomers, since such clusters are post-critical, but sufficiently small to reflect the nucleation event.\cite{FootNoteNuclLocation} For each MC trajectory, the identities of the bricks in the largest cluster were recorded at the last time step at which the largest cluster comprised 9 particles, and tallied over 60 independent simulations to give the frequency of brick types in the largest cluster. These frequencies were analysed at the temperatures at which self-assembly was `optimal' for both the fixed designed interactions and the DNA-specific interactions, \SI{323}{\kelvin} and \SI{318}{\kelvin}, respectively. These temperatures correspond to the lowest temperature at which structures self-assembled to large sizes without significant misassembly.

We first consider the frequencies of bricks in initial nucleation clusters for simulations of structures with assigned DNA sequences, where every pair of patches can interact with an interaction energy based on the SantaLucia thermodynamic model. Intuitively, we would expect that the stronger the bonding of a particle's patches is, the more likely it is for a particle to be found in the initial nucleus. Indeed, this is what is largely observed (Fig.~\refSubNB{(i)(a)}{fig-BBs-nucleationSites}): bricks with a higher average GC content (and hence stronger bonding) are more likely to be present in the critical nucleus. Intriguingly, it is not the bricks with the highest GC content that dominate; instead, nucleation tends to occur in regions where several neighbouring bricks have a high GC content. In other words, it appears that designing preferential nucleation pathways would require a careful analysis of not only the bonding strength of individual particles, but how they come together in the final structure, making it a more difficult task than it might first appear. We propose to investigate this interplay of factors more systematically in future work.

However, boundary bricks have a dominant effect as far as nucleation is concerned, and as soon as boundary bricks are added to a system, the small random variations in GC content that seem to determine the nucleation behaviour for the system with no BBs (Fig.~\refSubNB{(i)(a)}{fig-BBs-nucleationSites}) no longer play any significant role in determining a particle's nucleation propensity. For the structure with DNA-specific interactions and all 84 BBs present (Fig.~\refSubNB{(i)(b)}{fig-BBs-nucleationSites}), BBs are predominant in the initial nucleation cluster: the increased number of interactions per BB when compared to a `monomer' brick favours boundary bricks as preferred sites for nucleation to occur.

The same observation holds for the edge-BB and face-BB structures (Fig.~\refSubNB{(i)(c)}{fig-BBs-nucleationSites} and Fig.~\refSubNB{(i)(d)}{fig-BBs-nucleationSites}). These simulations demonstrate that it is possible to tune the nucleation site to different parts of the structure depending on where the BBs are located. The edge-BB structure nucleates essentially only at the edge BBs, and the face-BB structure nucleates only at the face BBs. The locations with the highest nucleation frequencies did contain the single BB among each subset with the highest GC content; however, beyond this, there was no significant correlation between nucleation frequency and GC content, perhaps indicating that the BB with the highest GC content is fastest to nucleate.

The nucleation location in structures with SantaLucia interactions is driven by both the location of BBs and the location of bricks with high GC content. BBs have a stronger proclivity for nucleation than monomers, and by selecting which bricks are bonded to others as BBs, we are able to control the nucleation site on the structure. For both the case without BBs and with BBs, bricks with higher GC content are involved in nucleation, though the exact effect of GC content on nucleation cannot be well understood from only one structure. Nevertheless, this finding could be a very useful tool in the rational design of self-assembly pathways of DNA brick structures.

We can investigate the underlying behaviour that is solely due to boundary bricks notwithstanding the effect of having varying interaction strengths across the system by considering the frequencies of bricks in initial nucleation clusters for the case of fixed designed interactions (Fig.~\refSub{ii}{fig-BBs-nucleationSites}). This system allows us to focus on only the size and geometry effects of BBs on nucleation, without the complication of non-uniform interaction strengths of DNA sequences. For the FDI structure with no BBs (Fig.~\refSubNB{(ii)(a)}{fig-BBs-nucleationSites}), nucleation appears to be dispersed throughout the volume of the structure, with bricks on the faces somewhat less likely to be involved in nucleation clusters. Since such bricks have non-interacting patches on the outside and therefore fewer possibilities for bonding, this behaviour is entirely consistent with expectations.

For the FDI structure with all 84 BBs present (Fig.~\refSubNB{(ii)(b)}{fig-BBs-nucleationSites}), nucleation is again largely confined to the faces and edges of the cube, where the BBs are located. Notably, nucleation essentially never occurs in the body of the structure. The presence of BBs, because they are larger units with more interaction sites per unit than regular bricks, causes nucleation to shift to the outer regions of the cube. This is further demonstrated by the FDI structures with only edge BBs (Fig.~\refSubNB{(ii)(c)}{fig-BBs-nucleationSites}) and only face BBs (Fig.~\refSubNB{(ii)(d)}{fig-BBs-nucleationSites}), where nucleation occurs primarily on the edges and faces, respectively, and is consistent with the behaviour seen in simulations with full sequence-dependent interactions: the nucleation propensities shown in Figs~\refSubNB{(i)(b--d)}{fig-BBs-nucleationSites} for full sequence-dependent interaction simulations largely correspond to those of Figs~\refSubNB{(ii)(b--d)}{fig-BBs-nucleationSites} of the analogous FDI simulations, demonstrating that the influence of GC content on the nucleation location is minimal as soon as boundary bricks are included. 

\subsection{Effect of BBs on the degree of assembly}
\begin{figure}[tbp]
\centering
\includegraphics{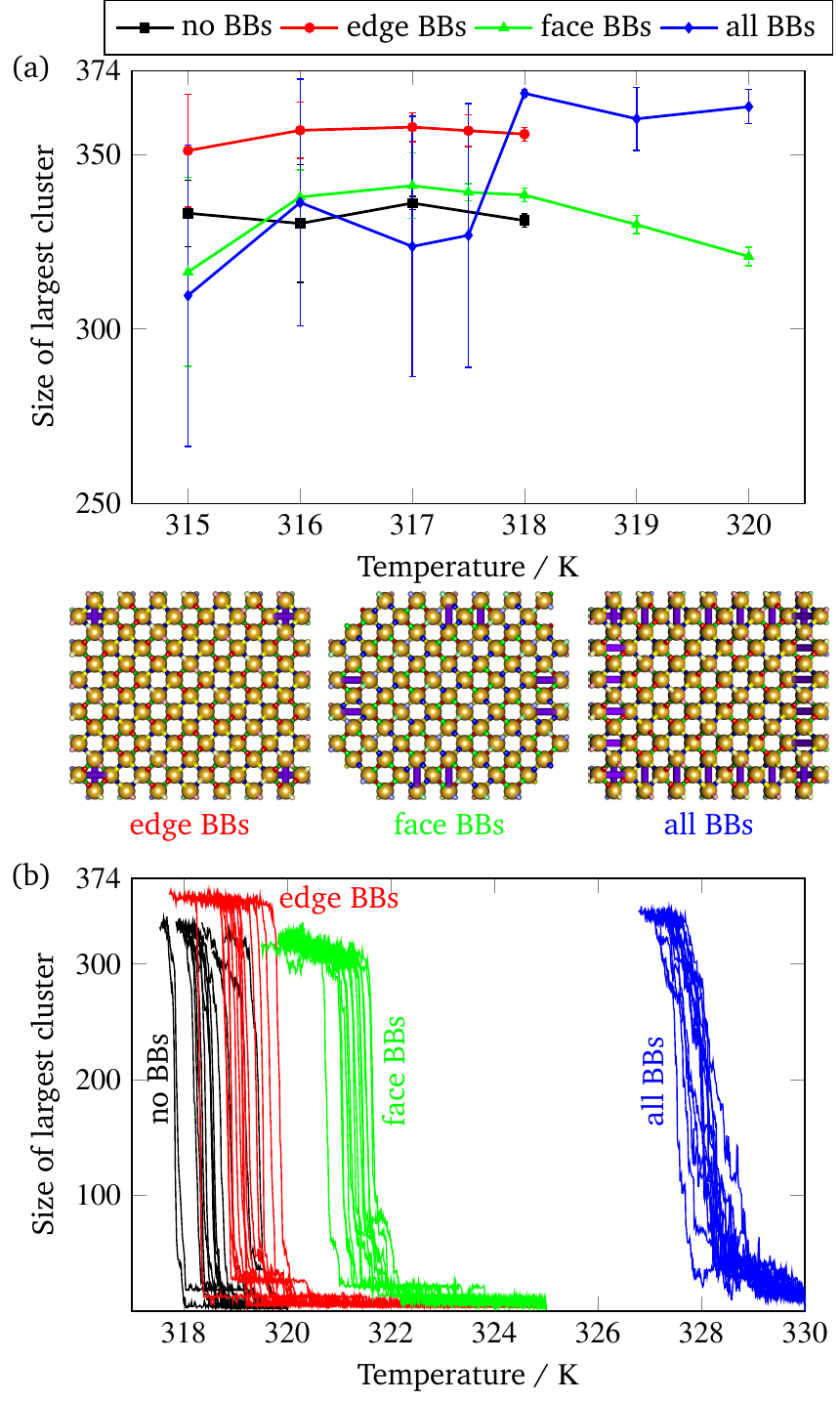}
\caption{(a) The size of the largest correctly assembled cluster in brute-force fixed-temperature simulations as a function of temperature for a system with full SantaLucia interaction parameters. Results averaged over 15 independent simulations at long times when the size has reached a plateau in the majority of simulations. We only report sizes for simulations where nucleation reliably occurred within \num{3e11} MC time steps. Large error bars indicate that clusters of various sizes have assembled, typically signalling that further growth is frustrated. Typical assembled structures corresponding to the three systems are shown below the graph in an orthographic projection. The orthographic projection emphasises the missing parts of the face-BB structure, but makes other missing bricks more difficult to see.  (b) The size of the largest correctly assembled cluster in the system as a function of temperature for simulations of systems with SantaLucia interactions. The cooling rate was \SI{1e-11}{\kelvin} per MC step.}
\label{fig-final-assembly-sizes}
\end{figure}

Although BBs lower the nucleation barrier for assembly, this is not necessarily beneficial to achieving successful self-assembly, as a lowered nucleation barrier may also lead to unwanted aggregation. While the all-BB system gets very close to growing to completion at \SI{318}{\kelvin} (Fig.~\refSub{a}{fig-nuc-rate}), none of the structures simulated quite reach the full size of the intended target of 374 particles at a temperature at which nucleation and designed growth occur. This is expected for a fixed temperature simulation, as partial assembly is entropically favoured,\cite{Jacobs2015b} and a temperature ramp is necessary to achieve full assembly, since at lower temperatures, the additional energetic stabilization drives the structure to assemble despite the entropic cost of full assembly.

\begin{figure*}[tbp]
\centering
\includegraphics{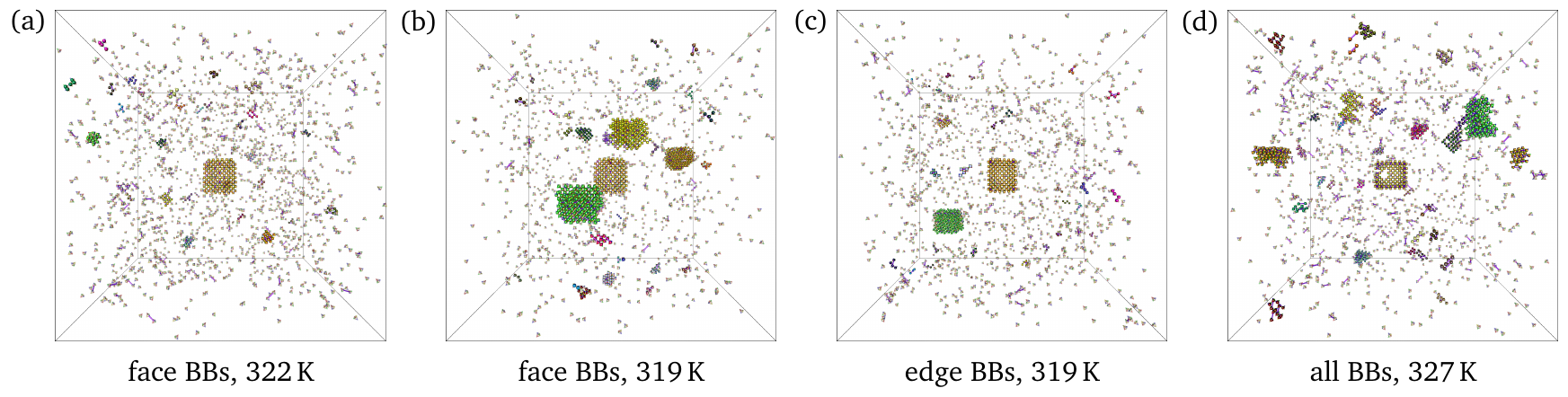}
\caption{Snapshots of typical configurations with multiple target structures in the simulation box, for systems with different boundary bricks at different temperatures (as labelled). These configurations are taken from constant-temperature simulations once the cluster sizes have stabilized over time. Clusters of particles bonded in the same way as in the target structures are shown in the same colour; different colours are used for clusters that are not connected to each other. Boundary bricks are connected by violet bonds, as in Fig.~\ref{fig-BB-ex-1}.}
\label{fig-multistruct-pics}
\end{figure*}

Since the entropic cost of binding a boundary brick is comparable to that of binding a non-brick monomer, but the degree of bonding can be greater, we expect that boundary bricks will stabilize the target structure in the sense that it can grow to a larger size even with a fixed-temperature growth protocol. In single-target simulations, the boundary bricks behave largely in the way we would expect them to: the all-BB structure grows to the largest final size in both fixed-temperature and gradually cooled set-ups (Fig.~\ref{fig-final-assembly-sizes}), as it has the largest number of boundary bricks to stabilize it.

However, the all-BB structure initially nucleates at very high temperatures, and it is only at approximately \SI{318}{\kelvin} that the target structure is nearly complete with few errors. Below this temperature, monomer nucleation is evidently too facile, which prevents successful assembly later on in the self-assembly process, as the probability of exactly aligning and forming all the right bonds to connect two larger clusters is prohibitively low. At low temperatures, self-assembly in simulations with all brick types becomes less and less favourable.

Interestingly, even though the edge-BB structures nucleate less rapidly than the face-BB structures, edge-BB structures had the second highest degree of final assembly across all optimal assembly temperatures (Fig.~\ref{fig-final-assembly-sizes}). This suggests that edge BBs are able to stabilize the final structure and bind bricks on the edge that would be entropically favoured to be unbound,\cite{Jacobs2015b} and are able to do so more effectively than face BBs. In Fig.~\refSub{a}{fig-final-assembly-sizes}, we show a typical example of a large correctly assembled cluster for each of the BB structures formed in constant-temperature simulations. In particular, the face-BB structure shown is missing all four edges. Of course, this is not wholly surprising, since the edge monomers typically have fewer bonds than the face monomers, and boundary bricks therefore play a much more significant role by comparison. This observation is supported by the fact that the face-BB structures, which only have normal monomer bricks on their edges, have nearly identical assembly sizes as the no-BB structures once this size has reached a plateau in constant-temperature simulations.

\subsection{Simulations of multiple target structures}

\begin{table}[tbp]
\caption{The number of large clusters and their average sizes formed in simulations with multiple copies of the target structure. Only clusters larger than a third of the final assembled size are considered to be large, and the `all clusters' column refers to clusters comprising more than 15 particles. All data averaged over 15 independent simulations after \num{e11} MC steps. The labels (a)--(d) correspond to Fig.~\ref{fig-multistruct-pics}.}\label{table-multiTarget}
\medskip
\centering
\begin{tabular}{lccccc} \toprule
\centering System & $T/\si{\kelvin}$ & \parbox{1.1cm}{\centering Number of all clusters} & \parbox{1.1cm}{\centering Number of large clusters} & \parbox{1.1cm}{\centering Mean size of large clusters} & \parbox{1.1cm}{\centering Median size of large clusters} \\ \midrule
(a) face BBs & 322 & 3.4 &  1.2  & 268 & 268    \\
(b) face BBs & 319 & 6.2 & 4.3 & 272 & 285   \\
(c) edge BBs & 319 & 4.5 & 2.5 & 314 & 314   \\
(d) all BBs & 327 & 11.5 & 3.5 & 204 & 195   \\ \bottomrule
 \end{tabular}
\end{table}
We have shown that both the nucleation behaviour of boundary bricks and their structure stabilization properties largely follow our na\"ive expectations: the more boundary bricks there are, the higher the nucleation temperature will be, and the more stabilized the target structure. Of course, in reality, more than a single target structure is normally assembled in solution; this complicates matters somewhat. In order to provide some insight into what the effect of boundary bricks might be in solution, we have therefore performed simulations in which several copies of the target structure are present.\cite{FootNoteGrandCanonical} In particular, we have simulated systems with 6 copies of each brick in the target structure at a density of $6/(100a)^3$. Results from these simulations are particularly interesting because unlike for single-target simulations, face-BB and edge-BB structures exhibit more facile self-assembly than the system with all possible BBs, as depicted in Fig.~\ref{fig-multistruct-pics}. The average cluster sizes corresponding to the conditions of Fig.~\ref{fig-multistruct-pics} are shown in Table~\ref{table-multiTarget}.

Of the systems studied, simulations with face-BB structures exhibit nucleation and growth over the largest range of temperatures. The temperature largely controls the number of large nuclei in the system: at \SI{322}{\kelvin}, only a single structure grows to an appreciable size (Fig.~\refSub{a}{fig-multistruct-pics}), whilst at \SI{320}{\kelvin}, up to 4 nearly complete structures self-assemble. At \SI{319}{\kelvin}, many simulations result in the successful self-assembly of roughly the same number of clusters (Fig.~\refSub{b}{fig-multistruct-pics}), but in a number of cases, these clusters merge incorrectly, and so the resulting structure can be classed as a kinetic aggregate. The behaviour of systems with edge-BB structures is similar, and, in keeping with the monomer results (see e.g.~Figs~\refSub{c}{fig-nuc-rate} and \refSub{b}{fig-final-assembly-sizes}), nucleation occurs at somewhat lower temperatures. However, aggregation is not shifted by the same amount in temperature, and so the range over which self-assembly occurs is narrower (roughly \SIrange{319}{320}{\kelvin}), and the number of large structures that grow successfully at this lower temperature is also smaller (typically only 2 at \SI{319}{\kelvin}, Fig.~\refSub{c}{fig-multistruct-pics}). As with single-target simulations, the protocol used for self-assembly is important: although clusters grow to medium sizes in multiple-target simulations after successful nucleation has occurred, the largest clusters can be made to grow essentially to completion if the system is subsequently cooled.

Finally, in simulations with all possible BBs, nucleation begins at very high temperatures ($\sim$\SI{330}{\kelvin}), consistent with single-target simulations. However, the clusters do not grow significantly at such high temperatures. As the temperature is decreased, the self-assembly process becomes very error-prone; whilst a single target structure typically grows much larger than the remaining structures, it often lacks the necessary components that have been used up to form other, smaller clusters already (see Table~\ref{table-multiTarget}), and so structures grow with large sections missing. For example, in Fig.~\refSub{d}{fig-multistruct-pics}, showing simulation results at \SI{327}{\kelvin}, several of the walls have nucleated separately from the rest of the target structure, making further growth very difficult. At \SI{326}{\kelvin}, several simulations resulted in the successful growth of a \emph{single} target structure (out of a possible 6 that could grow from the monomers), but in a similar number of simulations, no single target structure grew to a large size. Of course, since there are many monomers in solution, it is in some sense easier to assemble a single copy of the target structure than in single-target simulations; however, assembling multiple target structures simultaneously is difficult, since too many clusters nucleate and it is not straightforward then for them to meet in the correct geometry to form larger structures, and it appears Ostwald ripening is also not particularly fast. The choice of which boundary bricks to include when assembling a given target structure therefore appears to be a very important consideration in DNA brick self-assembly, and it appears from our simulations that opting for all possible boundary bricks is not the most favourable design choice.

One possible way of reducing the propensity for nucleation when many boundary bricks are included is to reduce their concentration relative to the remaining monomer bricks. To a first approximation, it is reasonable to assume that the chemical potential of a species appears in the same place in the hamiltonian as the binding energies. Increasing or decreasing the chemical potential of a species, for example by changing the species concentration, is thus effectively equivalent to shifting the strength of all the interactions of that species. We have run simulations of the all-BB system with only half the boundary bricks present. In keeping with expectations, the point at which nucleation occurs in brute-force simulations shifts to lower temperatures, and, with fewer clusters forming, multiple clusters can grow to larger sizes. Choosing an appropriate ratio of initial concentrations is therefore a further control parameter that can be tuned to improve assembly yields.

In our simulations, we observe both point defects as well as larger misbonded aggregates and missing features in the target structure. As far as we are aware, experiments on DNA brick systems have not thus far focussed on characterizing the nature of assembly errors in self-assembly, and indeed such defects may be rather difficult to probe experimentally; however, if we wish to ensure a faithful assembly of the complete target structure, this issue may be of great importance for the future of the field.

\section{Conclusions}

DNA brick structures have increasingly been studied over the last few years, since they provide a platform both for theoretical advancement in studying addressable self-assembly and for practical applications, such as creating nanostructures with nano-scale complexity for medicine, computing and nanoelectronics. In this work, we have extended a previously introduced patchy-particle model for DNA bricks to account for boundary bricks, which have been hypothesized to be an essential component of the experimental set-up for increasing the yield,\cite{Ke2012} but the effects of which had not previously been modelled.

It is important to bear in mind that our results correspond to a simple `toy model' of DNA bricks.  In reality, many effects that we have neglected may also be important, yet the system sizes involved are such that they make simulations with a more realistic potential intractable at present. However, the simplicity of our model suggests that our findings reflect the fundamental underlying physics of addressable self-assembly.

By simulating structures with varying placement of BBs in the canonical ensemble, we have shown that BBs located on the faces of the cubic target structure were primarily responsible for increasing the nucleation rate, whilst BBs located on the edges of the structure were primarily responsible for the stability of the final target structure. However, we have also found that structures that included BBs on all four possible faces were prone to misassembly, particularly in multiple-target simulations, as nucleation is too facile and multiple competing nuclei can grow and are subsequently unable to come together in the correct manner. This indicates that a strategy where all possible DNA strands that can be fused into boundary bricks actually are, as was done in previous experimental work, may not in fact be the optimal choice; a more careful consideration of the possible mechanisms of assembly and misassembly is warranted.

While the self-assembly pathways behave in fairly predictable ways in simulations where all patch--patch interactions are of the same strength, further complications arise when DNA sequences are taken into account, since the dominant nucleation locations depend on the strengths of the nearby interactions. We have briefly investigated this effect by examining the GC content of the brick structures, and we found that it was regions with a higher than average GC content that were most likely to nucleate first, rather than necessarily single bricks with an especially high GC content. It would be particularly interesting to investigate this behaviour further and determine whether any simple rules that govern the nucleation location as a function of interaction strength can be identified. However, although the GC content seems to play a significant role for structures without boundary bricks, as soon as BBs are included, the nucleation location is almost completely dominated by the BBs: in structures where BBs are localized to the edges, the edges were involved in nucleation, whilst in structures where BBs are localized to the faces, nucleation occurred primarily on the faces.

However, we have shown that BBs affect more than just nucleation. Since they entail the formation of more bonds, the bonding of a BB to the growing structure results in a more favourable enthalpic contribution to the free energy than a monomer brick would give, whilst the loss of entropy is only marginally more disfavourable. This means that target structures can grow larger at a given temperature than they would for a system without boundary bricks. In particular, boundary bricks can stabilize any `fine structure' on the surface of the target structure, which could be especially important for those practical applications for which the assembled structure must be as perfect as possible.

Although boundary bricks do allow us to construct structures that are more `perfect' in their final assembled state, including them can be something of a double-edged sword, since they not only stabilize the final structure, but are also easier to nucleate, which means they are more prone to misassembly and aggregation. This may to a significant extent negate the benefit of brick self-assembly being a nucleation-initiated process. In practice, it might be necessary to balance the two effects. For example, it may be possible to increase the yield by keeping the concentration of BBs lower than that of the monomer bricks or keeping the temperature higher for longer in order to keep nucleation a sufficiently rare event. It could also be possible to make the average bonding strength in boundary bricks weaker than that of the remaining monomer bricks, reducing the propensity for premature nucleation, whilst still allowing a degree of stabilization of the final structure. However, we have found that in our simulations, even when multiple structures were allowed to compete with one another in the same simulation box, there was a range of temperatures at which nucleation was rate limiting, but nevertheless sufficiently common for multiple target structures to grow essentially to completion even when boundary bricks were included.

Including all possible bricks was not particularly advantageous for assembly in multiple-target simulations, and including only face or only edge BBs resulted in self-assembly that was much less error prone. We envisage that a careful consideration of the types of boundary brick to include to maximize the yield and the quality of the target structures will be particularly important when looking at more complex structures than the ones we considered here. We have only looked at a cubic target structure in this work, as such a system is easiest to study systematically. When target structures include a complex array of peaks and troughs, the choice of the types of boundary brick which will stabilize the target structure whilst minimizing misassembly is considerably less straightforward. Having a clear design strategy is even more important for such systems, but intuition alone may not be enough; indeed, a simple simulation with our coarse-grained potential may well provide a convenient design tool for this purpose. Our simple coarse-grained potential may provide a useful first approximation when faced with a realistic design problem involving DNA bricks, and we hope that our work will provide useful insight to experimentalists interested in the practical applications of such systems.

\section*{Acknowledgements}
We thank Martin Sajfutdinow, David M.~Smith, William M.~Jacobs and Thomas E.~Ouldridge for helpful discussions. This work was supported by the Engineering and Physical Sciences Research Council [Programme Grant EP/I001352/1]. HKWS acknowledges support from the Winston Churchill Foundation of the United States. Research carried out in part at the Center for Functional Nanomaterials, Brookhaven National Laboratory, which is supported by the US Department of Energy, Office of Basic Energy Sciences, under Contract No.~DE-SC0012704.

Supporting data are available at the University of Cambridge Data Repository, \href{https://doi.org/10.17863/cam.7049}{doi:10.17863/cam.7049}.

\footnotesize{
\providecommand*{\mcitethebibliography}{\thebibliography}
\csname @ifundefined\endcsname{endmcitethebibliography}
{\let\endmcitethebibliography\endthebibliography}{}

}


\begin{mcitethebibliography}{42}
\bibitem[FootNotePolyT()]{FootNotePolyT}
A patch can be passivated by having a `poly-T sequence', which is a DNA
  sequence of 8 consecutive thymines. No other patches in the system have a
  sequence complementary to a poly-T sequence, which therefore minimizes the
  patch's chance of misbonding.\relax
\mciteBstWouldAddEndPunctfalse
\mciteSetBstMidEndSepPunct{\mcitedefaultmidpunct}
{}{\mcitedefaultseppunct}\relax
\EndOfBibitem
\bibitem[FootNoteBoundaryCount()]{FootNoteBoundaryCount}
In order to enable us to compare to the same target structure size regardless
  of the number of boundary bricks in the system, we count each `monomer' in a
  boundary brick independently when computing the sizes of correctly assembled
  clusters.\relax
\mciteBstWouldAddEndPunctfalse
\mciteSetBstMidEndSepPunct{\mcitedefaultmidpunct}
{}{\mcitedefaultseppunct}\relax
\EndOfBibitem
\bibitem[FootNoteFaceBBs()]{FootNoteFaceBBs}
The boundary bricks coloured in blue in Fig.~\ref{fig-BBs-allTypes} are, of
  course, also on the faces of the structure. We chose to consider the central
  face boundary bricks separately from the remaining face boundary bricks in
  order to ensure that the number of face boundary bricks was the same as the
  number of edge boundary bricks. The term `face-BB structure' refers solely to
  these central face boundary bricks, which are shown in green in
  Fig.~\ref{fig-BBs-allTypes}.\relax
\mciteBstWouldAddEndPunctfalse
\mciteSetBstMidEndSepPunct{\mcitedefaultmidpunct}
{}{\mcitedefaultseppunct}\relax
\EndOfBibitem
\bibitem[FootNoteNuclLocation()]{FootNoteNuclLocation}
The same analysis was performed for clusters comprising 8, 10, 15, and 20
  bricks, with similar frequencies.\relax
\mciteBstWouldAddEndPunctfalse
\mciteSetBstMidEndSepPunct{\mcitedefaultmidpunct}
{}{\mcitedefaultseppunct}\relax
\EndOfBibitem
\end{mcitethebibliography}
\end{document}